\documentclass[12pt]{article}
\let\section=\subsection     \let\subsection=\subsubsection                %%
\usepackage{graphicx}
\input{psfig}
\def\Journal#1#2#3#4{{#1} {\bf #2} (#3) #4}

\def\PRL{ Phys. Rev. Lett.}
\def\PRD{{ Phys. Rev.} D}

\begin{document}
\begin{center}
   {\large \bf GLUEBALLS, HYBRID AND EXOTIC MESONS}\\[5mm]
   C.~MICHAEL \\[5mm]
   {\small \it Theoretical Physics Division, Dept. of Math. Sci.,\\
 University of Liverpool, Liverpool, L69 2BX, U.K. \\[8mm] } 
\end{center}

\begin{abstract}\noindent
 We review lattice QCD results for glueballs
(including  a discussion of mixing with scalar mesons), hybrid mesons
and exotic mesons (such  as $B_s B_s$ molecules). 
\end{abstract}

\section{Introduction}

The most systematic approach to non-perturbative QCD is via lattice
techniques. 
  Lattice QCD needs as input the quark masses and an overall scale
(conventionally  given by $\Lambda_{QCD}$). Then any Green function can
be evaluated by taking an average of suitable combinations of the
lattice fields in the vacuum samples. This allows masses to be studied 
easily and matrix elements (particularly those of weak or
electromagnetic currents)  can be extracted straightforwardly.
  Unlike experiment, lattice QCD can vary the quark masses and can also 
explore different boundary conditions and sources. This allows a wide
range of  studies which can be used to diagnose the health of
phenomenological models as well as casting light on experimental data.

One limitation of the  lattice approach  to QCD is  in exploring
hadronic decays because the  lattice, using Euclidean
time, has no concept of asymptotic  states. One feasible strategy is to
evaluate the mixing between states of the same  energy - so giving some
information on on-shell hadronic decay amplitudes.

 For comparison with models and for ease of computation, the special 
case  of infinitely heavy sea quarks (namely neglect of quark effects in
the vacuum:  the  quenched approximation) is often used. We shall also 
present results from including sea quark effects - usually two flavours
of  degenerate sea quark of mass equivalent to strange quarks or
heavier.

 The quark model gives a good overall description of hadronic spectra.
Here I  will discuss lattice results for states which go beyond the
quark model:  glueballs, exotic mesons and hybrid mesons.

\section{Glueballs}

Glueballs are defined to be hadronic states made primarily from gluons.
The full non-perturbative gluonic interaction is included in quenched
QCD.  In the quenched approximation, there is no mixing between such
glueballs  and quark - antiquark mesons. A study of the glueball
spectrum in quenched QCD  is thus of great value. This will allow
experimental searches to be  guided as well as providing calibration for
models of glueballs. A non-zero glueball mass in quenched QCD is the 
``mass-gap'' of QCD. To prove this rigourously is one of the major
challenges  of our times. Here we will explore the situation using
computational techniques.

In principle, lattice QCD can study the meson spectrum as the sea quark
mass  is decreased towards experimental values. This will allow the
unambiguous glueball states in the quenched approximation to be tracked
as the sea quark effects are increased.  It may indeed turn out that no
meson in the physical spectrum is primarily a glueball - all states are 
mixtures of glue,  $q \bar{q}$, $q \bar{q} q \bar{q}$, etc. We shall
later discuss  lattice results on the mixing of glueballs and scalar
mesons (ie $q \bar{q}$ states). 

In lattice studies, dimensionless ratios of  quantities are obtained. To
explore the glueball masses, it is appropriate to combine  them with
another very accurately measured quantity to have a dimensionless 
observable. Since the potential between static quarks is very accurately
measured from the lattice, it is now conventional to use $r_0$ for this
comparison.  Here $r_0$ is implicitly defined by $r^2 dV(r)/dr = 1.65$
at $r=r_0$ where $V(r)$ is  the potential energy between static quarks
which is easy to determine accurately  on the lattice.  Conventionally 
$r_0 \approx 0.5$ fm.

 Theoretical analysis  indicates that for  Wilson's discretisation of
the gauge fields in the quenched approximation,  the dimensionless ratio
$mr_0$ will differ from the continuum  limit value by corrections of
order $a^2$.  Thus in fig.~1 the mass of the $J^{PC}$=$0^{++}$  glueball
is plotted versus the lattice spacing $a^2$. The straight line then
shows the continuum limit obtained  by extrapolating to $a=0$. As can be
seen, there is essentially no need for data  at even smaller $a$-values
to further fix the continuum value. The value shown  corresponds to
$m(0^{++})r_0=4.33(5)$.  Since several lattice
groups~\cite{DForc,MTgl,ukqcd,gf11} have measured these  quantities, it
is reassuring to see that the purely lattice observables are in 
excellent agreement. The publicised difference of quoted $m(0^{++})$
from  UKQCD~\cite{ukqcd} and GF11~\cite{gf11} comes entirely from
relating quenched lattice  measurements to values in GeV.

\begin{figure}[bt] 
\vspace{7cm} %
\includegraphics{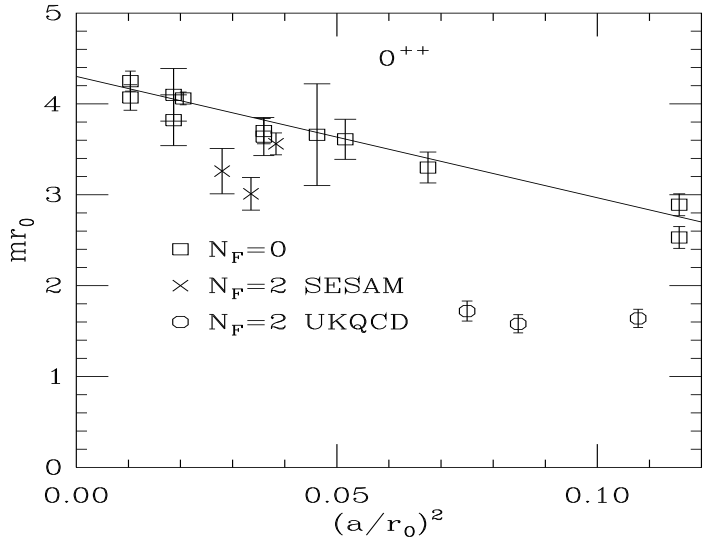}
 \caption{ The value of mass of the  $J^{PC}=0^{++}$  glueball state
from quenched data ($N_F=0$){\protect\cite{DForc,MTgl,ukqcd,gf11}}
in units of $r_0$ where $r_0 \approx 0.5$ fm. The straight line  shows a
 fit describing the  approach to the continuum limit as $a \to 0$.
 Results~{\protect\cite{sesam,bali,lat99}} with $N_F=2$ flavours of sea
quarks are also shown.
   }
\end{figure}

In the quenched approximation, different hadronic observables differ
from experiment  by factors of up to 10\%. Thus using one quantity or
another to set the scale, gives an overall systematic error.  Here I
choose to set the scale by taking the conventional value of the string
tension, $\sqrt{\sigma}=0.44$ GeV, which then corresponds to
$r_0^{-1}=373$ MeV. An overall systematic error of 10\% is then to be
included to any  extracted mass. This yields $m(0^{++})=1611(30)(160)$
MeV where the second error is the
systematic  scale error. Note that this is the  glueball mass in the
quenched approximation -  in the real world significant mixing with $q
\bar{q}$ states could modify this value substantially.

In the Wilson approach, the next lightest glueballs
are~\cite{MTgl,ukqcd} the tensor $m(2^{++})r_0=6.0(6)$  (resulting in  
$m(2^{++})=2232(220)(220)$ MeV) and the pseudoscalar $m(2^{++})r_0=
6.0(1.0)$. Although the Wilson discretisation provides a definitive
study of the lightest ($0^{++}$)  glueball in the continuum limit, other
methods are competitive for the determination of the mass  of heavier
glueballs.  Namely, using an improved gauge discretisation which has 
even smaller discretisation errors than the $a^2$ dependence of the
Wilson discretisation,  so allowing a relatively coarse lattice spacing
$a$ to be used. To extract mass values, one has to explore the time
dependence of correlators and for this reason,  it is optimum to use a
relatively small time lattice spacing. Thus an asymmetric  lattice
spacing is most appropriate.  The  results~\cite{mpglue}  are shown in
fig.~2 and for low lying states are that $m(0^{++})r_0=4.21(11)(4)$, 
$m(2^{++})r_0=5.85(2)(6)$, $m(0^{-+})=6.33(7)(6)$ and
$m(1^{+-})r_0=7.18(4)(7)$.  Another recent study~\cite{nrw} has used an
improved discretisation based on the perfect action  approach (without a
space-time asymmetry) and obtains  results consistent with earlier work.

One signal of great interest would be  a glueball with $J^{PC}$ not
allowed for $q \bar{q}$ - a spin-exotic glueball or {\em oddball} -
since it would  not mix with $q \bar{q}$ states. These states are
found~\cite{MTgl,ukqcd,mpglue} to be  high lying: considerably above
$2m(0^{++})$. Thus they are  likely to be in a region very difficult to
access unambiguously by experiment.

 Within the quenched approximation, the glueball states are unmixed 
with $q \bar{q},\ q \bar{q} q \bar{q}$, etc. Furthermore, the  $q
\bar{q}$ states have degenerate flavour singlet and non-singlet states
in the quenched approximation.  Once quark loops are allowed in the
vacuum, for the favour-singlet states of any given $J^{PC}$,  there will
be mixing between the  $s \bar{s}$ state, the  $u \bar{u}+d \bar{d}$
state  and the glueball. 
 One way to explore this is to measure directly the scalar  mass
eigenstates in a study with $N_f=2$ flavours of sea-quark.
 Most studies show no significant change of the glueball spectrum as
dynamical quark effects are added - but  the sea quark masses used are
still rather large~\cite{sesam,bali}. A recent study~\cite{lat99},
however,  does find evidence for a reduced mass, albeit with a rather
large lattice spacing,  see fig.~1.
 This effect could be due to mixing of scalar mesons and glueballs, as
we discuss  below, or might just be a sign of an enhanced order $a^2$ 
correction at the  relatively large lattice spacing used. 

Let us now discus the mixing of the scalar glueball and scalar mesons.
The  mass spectrum of $q \bar{q}$ states has been determined on a 
quenched lattice and  the scalar mesons  are found to lie somewhat
lighter than the tensor states~\cite{livhyb}. These  $2^{++}$ mesons are
experimentally almost unmixed and  so  will be quite close to the
quenched mass determination. This suggests  that the quenched scalar
masses from the lattice are at around 1.2 GeV and 1.5 GeV (for $n
\bar{n}$ and $s \bar{s}$ respectively). An independent
study~\cite{weinss,weinssg}  suggests that the scalar $s \bar{s}$ state
is about 200 MeV lighter than the glueball which is a broadly compatible
conclusion.  Thus the glueball, at around 1.6 GeV, lies heavier than the
lightest $q\bar{q}$ scalar states. This information can then be combined
with mixing strengths to give the resulting scalar spectrum. 

It is possible to measure the mixing strength on a quenched lattice even
though no mixing actually occurs. On a rather coarse lattice ($a^{-1} 
\approx 1.2$ GeV), two groups  have attempted this~\cite{weinssg,lat99}.
Their results expressed as the mixing for two degenerate quarks of mass
around the strange quark mass  are similar, namely $E \approx 0.36$
GeV~\cite{weinssg} and 0.44 GeV~\cite{lat99}. From this evaluation of the
mixing strength, one can use  a mass matrix to estimate the mass shift
induced in the glueball and  scalar meson.
 The relevant mass matrix   is (in a glueball, $q \bar{q}$ basis in GeV
units):

\begin{math} 
 \left( \begin{array}[h]{cc}
    1.1  &    0.4 \\
    0.4  &   1.6 \\
\end{array} \right)
\end{math}

\noindent which would give a downward shift of the glueball mass by
20\%. 
 This is in qualitative agreement with our direct determination with 
$N_f=2$ flavours of sea-quark that the lightest scalar mass is reduced 
significantly at this lattice spacing as shown in fig.~1.

 Note that at this coarse lattice spacing the quenched glueball mass is 
reduced (see fig.~1) below the canonical value of 1.6 GeV. Thus a study 
at smaller lattice spacing is needed. An exploratory attempt to
extrapolate to  the continuum~\cite{weinssg} gave a very small mixing of
86(64) MeV, while the  other determination~\cite{lat99} uses clover
improvement so order $a$ effects in the extrapolation to the continuum
are suppressed and one would not expect a significant decrease in going
to  the continuum limit. 
 What this discussion shows is that precision studies of the mixing on 
a lattice have not yet been achieved.

\begin{figure}[t]
\psfig{figure=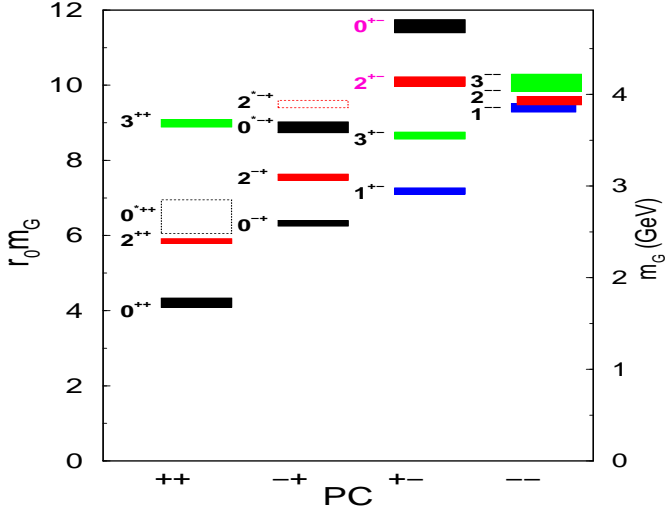,height=7cm,width=9cm}
\vspace{-0.5cm}
 \caption{ The continuum glueball spectrum{\protect\cite{mpglue}}. 
 }
 \label{gbr0}
\end{figure}

As well as this mixing of the glueball with $q \bar{q}$ states, there
will be  mixing  with $q \bar{q} q \bar{q}$ states which will be
responsible for the  hadronic decays. A first attempt to study
this~\cite{gdecay} yields an estimated width for decay to two
pseudoscalar mesons from the scalar glueball of order 100 MeV.  A more
realistic study  would involve taking account of mixing with the $n
\bar{n}$ and $s \bar{s}$ scalar mesons as  well.

\section{Exotic states}

 By exotic state we mean any state which is not dominantly a $q \bar{q}$
 or $qqq$ state. Such examples have been known for a long time: the
deuteron  is a proton-neutron molecule for example. It is very weakly
bound (2 MeV)  and is quite extended. Similar molecular states involving
two mesons have been conjectured. 

 One case which is relatively easy to study is the $BB$ system,
idealised as two  static quarks and two light quarks. Then a potential
as a function of the separation  $R$ between the static quarks can be
determined.  Because the static quark spin is irrelevant, the states can
be classified by the light quark spin and  isospin.  Lattice
results~\cite{cmpp}  (using a light quark mass close to strange) have
been obtained for the potential energy for $I_q=0,1$ and $S_q=0,1$. For
very  heavy quarks, a potential below $2M_B$  will imply binding of the
${BB}$ molecules with these quantum numbers and $L=0$. For the
physically relevant case  of $b$ quarks of around 5 GeV, the kinetic
energy will not be negligible and the binding energy of the ${ BB}$
molecular states is less  clear cut. One way to estimate the kinetic
energy for the ${ BB}$ case with reduced mass circa 2.5 GeV is to use
analytic approximations to the  potentials found. For example the
$I_q,S_q$=(0,0) case shows a deep  binding at $R=0$ which  can be
approximated as a Coulomb potential of $-0.1/R$ in GeV units. This will
give a di-meson binding energy of only 10 MeV.  For the other
interesting case, $(I_q,S_q)$=(0,1), a  harmonic oscillator potential in
the radial coordinate of form $-0.04[ 1- (r-3)^2/4]$ in GeV units leads
to a kinetic energy  which completely cancels the potential energy
minimum, leaving zero  binding. This harmonic oscillator approximation
lies above the estimate of  the potential, so again we expect weak
binding of the di-meson system.

 Because of these very small values for the di-meson binding energies, 
one needs to retain corrections to the heavy quark approximation to 
make more definite predictions, since these corrections are known to 
be of magnitude 46 MeV from the $B$, $B^*$ splitting. It will also be 
necessary to extrapolate the  light quark mass from strange to 
the lighter $u,\ d$ values to make more definite predictions 
about the binding of $BB$ molecules.

Models for the binding of two $B$ mesons involve, as in the case of the
deuteron,  pion exchange. The lattice study~\cite{cmpp} is able to make a
quantitative comparison of lattice pion  exchange with the data
described above and excellent agreement is obtained at larger $R$ 
values, as expected.

 \section{Hybrid Mesons}

 By hybrid meson, I mean a meson in which the gluonic degrees of freedom
are  excited non-trivially. 
 I first discuss hybrid mesons with static heavy quarks where the
description  can be thought of as an excited colour string. I then
summarise the situation  concerning light quark hybrid mesons.

Consider $Q \bar{Q}$ states with static quarks  in which the gluonic
contribution may be excited. We  classify the gluonic fields according
to the symmetries of the system.  This discussion is very similar to the
description of electron wave functions in  diatomic molecules. The
symmetries are  (i) rotation around the separation axis $z$ with
representations labelled by $J_z$ (ii) CP with representations labelled
by $g(+)$ and $u(-)$ and (iii) C$\cal{R}$. Here  C interchanges $Q$ and
$\bar{Q}$, P is parity and $\cal{R}$ is a rotation  of $180^0$ about the
mid-point around the $y$ axis. The C$\cal{R}$ operation is only relevant
 to classify states with $J_z=0$. The convention is to label states of
$J_z=0,1,2$ by $ \Sigma, \Pi, \Delta$  respectively. The ground state
($\Sigma^+-g$) will have $J_z=0$ and $CP=+$.

 The exploration of the energy levels  of other representations has a
long history in lattice studies~\cite{liv,pm}. The first excited state
is found  to be the $\Pi_u$.  This can be visualised  as the symmetry of
a string bowed out in the $x$ direction minus the same  deflection in
the $-x$ direction (plus another component of  the two-dimensional
representation with the transverse direction $x$ replaced by $y$),
corresponding to flux  states from a lattice  operator which is the
difference of U-shaped paths from quark to antiquark of the form $\,
\sqcap - \sqcup$.

Recent lattice studies~\cite{jkm}  have used an asymmetric space/time
spacing which enables excited states to be  determined comprehensively.
 These results confirm the finding that 
the $\Pi_u$ excitation is the lowest lying and hence of most relevance 
to spectroscopy.

 From the potential corresponding to these excited gluonic states, one
can  determine the spectrum of hybrid quarkonia using the Schr\"odinger
equation in the Born-Oppenheimer approximation.  This approximation will
be good if the heavy quarks move very little in the  time it takes for
the potential between them to become established. More  quantitatively,
we require that the potential energy of gluonic excitation is much
larger than the typical energy of orbital or radial excitation.  This is
indeed the case~\cite{liv}, especially for $b$ quarks. Another nice
feature of this approach is that the  self energy of the static sources
cancels in the energy difference between this  hybrid state and the
$Q \bar{Q}$ states. Thus the lattice approach gives directly the
excitation energy  of each gluonic excitation.

  The $\Pi_u$ symmetry state corresponds to  excitations of the gluonic
field in quarkonium called magnetic (with $L^{PC}=1^{+-}$) and
pseudo-electric (with $1^{-+}$) in contrast to the usual  P-wave orbital
excitation which has $L^{PC}=1^{--}$. Thus we expect different quantum
number assignments from those of the gluonic ground state. Indeed
combining with the heavy quark spins, we get a degenerate  set of 8
states with    $J^{PC}=1^{--}$, $ 0^{-+}$, $ 1^{-+}$, $ 2^{-+}$ and  
$1^{++},\ 0^{+-},\ 1^{+-},\ 2^{+-}$  respectively. Note that of these, 
$J^{PC}=  1^{-+},\ 0^{+-}$ and   $2^{+-}$  are spin-exotic and hence
will not mix with $Q\bar{Q}$ states. They thus form a very attractive
goal for experimental searches for hybrid  mesons.

 The eightfold degeneracy of the static approach will be broken by 
various corrections. As an example, one of the eight degenerate  hybrid
states is a pseudoscalar with the heavy quarks in a spin triplet.  This
has the same overall quantum numbers as the S-wave  $Q \bar{Q}$ state
($\eta_b$) which, however, has the heavy quarks in a spin singlet. So
any  mixing between these states must be mediated by spin dependent
interactions.  These spin dependent interactions will be smaller for
heavier quarks. It is  of interest to establish the strength of these
effects for $b$ and $c$ quarks. Another topic of interest is the
splitting  between the spin exotic hybrids which will come from the
different  energies  of the magnetic and pseudo-electric gluonic
excitations.

 One way to go beyond the static approach is to use the NRQCD
approximation which then enables  the spin dependent effects to be
explored.  One study~\cite{jkm} finds that the  $L^{PC}=1^{+-}$ and
$1^{-+}$ excitations  have no statistically significant splitting 
although the $1^{+-}$  excitation does lie a little lighter. This would
imply, after adding in heavy quark spin, that  the $J^{PC}=1^{-+}$
hybrid was the lightest spin exotic. Also a relatively large spin
splitting was found~\cite{cppacs} among the triplet states considering,
however,   only
 magnetic gluonic excitations.

 Confirmation of the ordering of the spin exotic states also comes from
 lattice studies with propagating quarks~\cite{livhyb,milc,sesamhyb}
which  are able to measure masses for all 8 states. We  discuss this
evidence in more detail below.

 Within the quenched approximation,  the lattice evidence  for
$b\bar{b}$ quarks points to a  lightest hybrid spin exotic with
$J^{PC}=1^{-+}$ at an energy given by $(m_H-m_{2S})r_0$ =1.8 (static
potential~\cite{pm}); 1.9 (static potential~\cite{jkm},
NRQCD~\cite{cppacs}); 2.0 (NRQCD~\cite{jkm}). These results can be
summarised as       $(m_H-m_{2S})r_0=1.9 \pm 0.1$.
 Using the experimental mass of the $\Upsilon(2S)$, this implies that
the lightest spin exotic  hybrid is at $m_H=10.73(7)$ GeV including a
10\% scale error.  Above this energy there will be many more hybrid 
states, many of which will be spin exotic. A  discussion of hybrid decay 
channels has been given~\cite{hf8}.

 The  excited gluonic static potential has also been determined
including sea quarks  ($N_f=2$ flavours) and no significant difference
is seen~\cite{bali}. Thus the quenched estimates given above are not
superseded. 
 %- see fig.~\ref{balif}

 I now  focus on lattice results for hybrid mesons made from light
quarks using fully relativistic propagating quarks.  There will be no
mixing with $q \bar{q}$ mesons for  spin-exotic hybrid mesons  and these
are of special interest. The first study of this area was by the  UKQCD
Collaboration~\cite{livhyb} who used operators motivated by the  heavy
quark studies referred to above to study all 8 $J^{PC}$ values coming
from $L^{PC}=1^{+-}$ and $1^{-+}$ excitations. The  resulting mass
spectrum  gives the $J^{PC}=1^{-+}$ state as the lightest spin-exotic
state. Taking account of the systematic scale errors in the lattice
determination, a  mass of 2000(200) MeV is quoted for this hybrid meson
with $s \bar{s}$ light quarks. Although not directly measured, the
corresponding light quark hybrid meson would be expected to be around
120 MeV lighter.

A second lattice group has also evaluated hybrid meson spectra with
propagating quarks from quenched lattices. They obtain~\cite{milc}
masses of the $1^{-+}$ state with statistical and various systematic
errors of  1970(90)(300) MeV, 2170(80)(100)(100) MeV and 4390(80)(200)
MeV for $n \bar{n}$,  $s \bar{s}$ and $c \bar{c}$ quarks respectively.
For the  $0^{+-}$ spin-exotic state they have a noisier signal but
evidence that it is heavier. They also explore mixing matrix elements
between spin-exotic hybrid  states and 4 quark operators. 

 A first attempt has been made~\cite{sesamhyb} to determine the hybrid
meson spectrum using  full QCD. The sea quarks used have several
different masses and an extrapolation  is made to the limit of physical
sea quark masses, yielding a mass of 1.9(2) GeV for the lightest 
spin-exotic hybrid meson, which again is found to be the $1^{-+}$. In
principle this  calculation should take account of sea quark effects
such as the mixing  between such a hybrid meson and $q \bar{q} q
\bar{q}$ states such as $\eta \pi$, although it is possible that the sea
quark  masses used are not light enough to explore these features.

The three independent lattice calculations of the light hybrid spectrum
are  in good agreement with each other. They imply that the natural
energy  range for spin-exotic hybrid mesons is around 1.9 GeV. The
$J^{PC}=1^{-+}$  state is found to be lightest. It is not easy to
reconcile these lattice results  with experimental
indications~\cite{expt} for resonances at 1.4 GeV and 1.6 GeV,
especially the  lower mass value.  Mixing  with  $q \bar{q} q \bar{q}$
states such as $\eta \pi$ is not included for realistic quark masses in
the  lattice calculations. This can be interpreted, dependent on one's
viewpoint,  as either that the lattice calculations  are incomplete or
as an indication that the experimental states may have an  important
meson-meson component in them.

 \section{Conclusions}

 Quenched lattice QCD is well understood and accurate predictions in the
continuum limit  are increasingly becoming available. The lightest
glueball  is scalar with mass  $m(0^{++})=1611(30)(160)$ MeV where the
second error is an overall scale error. The excited glueball spectrum is
known too. The quenched approximation  also gives information on
quark-antiquark scalar mesons and their mixing with glueballs. This
determination of the mixing in the quenched approximation  also sheds
light on results for the  spectrum directly  in full QCD where the
mixing will be enabled. There is also some lattice information  on the
hadronic decay  amplitudes of glueballs.

 Evidence exists for a  possible $B_s B_s$ molecular state.

 For hybrid mesons, there will be no mixing with $q \bar{q}$ for 
spin-exotic states and these are the most useful predictions. The
$J^{PC}=1^{-+}$ state is expected at 10.73(7) GeV for $b$ quarks,
 2.0(2) GeV for $s$ quarks and 1.9(2) GeV 
for $u,\ d$ quarks. Mixing of spin-exotic hybrids with
$q\bar{q}q\bar{q}$ or equivalently with meson-meson  is  allowed and
will modify the  predictions from the quenched approximation.

%% Placing of figures:
%% If you have problems to place figures appropriately in the text with the
%% figure-environment, use this construction: 
%\begin{center}
%   \includegraphics[width=6cm,height=5cm,angle=-90]{figure_1.eps}\\
%   \parbox{14cm}
%	{\centerline{\footnotesize 
%	Fig.~1: Spectral function (left) of
%	the $\rho$-meson as a function \dots}}
%\end{center}


\begin{thebibliography}{99}
\itemsep=0cm

%glueball

\bibitem{DForc} P. De Forcrand et al., { Phys.\ Lett.} {\bf B152}, 
 (1985) 107.

\bibitem{MTgl} C. Michael and M. Teper, { Nucl.\ Phys.} {\bf B314}
(1989) 347. 

\bibitem{ukqcd}  UKQCD collaboration, G. Bali, et al.,
{ Phys.\ Lett.} {\bf B309} (1993) 378. 

\bibitem{gf11} H. Chen et al.,
{ Nucl.\ Phys. B (Proc.\ Suppl.)} {\bf 34} (1994) 357; 
A. Vaccarino and D. Weingarten, \Journal{\PRD}{60}{1999}{114501}.

\bibitem{sesam}G. Bali et al., 
{ Nucl.\ Phys. B (Proc.\ Suppl.)} {\bf 63} (1998) 209.

\bibitem{lat99}    C. Michael, M. S. Foster and C. McNeile,
{ Nucl. Phys. B (Proc. Suppl.)} {\bf 83-84} (2000) 185;
 C. McNeile and C. Michael, LTH487, hep-lat/0010019.

\bibitem{mpglue} C. Morningstar and M. Peardon, { Phys. Rev.} {\bf
D56} (1997) 4043; { ibid.}, {\bf D60} (1999) 034509.


\bibitem{nrw} F. Niedermeyer, P. R\"ufenacht and U. Wenger,
hep-lat/0007007.

\bibitem{livhyb} UKQCD Collaboration,
  P. Lacock, C. Michael, P. Boyle  and P. Rowland, 
{ Phys.\ Rev.} {\bf D54} (1996)  6997; 
{ Phys.\ Lett.} {\bf B401} (1997)  308; 
 { Nucl. Phys. B (Proc. Suppl.)} {\bf 63} (1998)  203.

\bibitem{weinss} W. Lee and D. Weingarten, { Nucl. Phys. B (Proc.
Suppl)} {\bf 53} (1997) 236; { Nucl. Phys. B (Proc.  
Suppl)} {\bf 73} (1999) 249.

\bibitem{weinssg}  W. Lee and D. Weingarten, { Nucl. Phys. B (Proc.
Suppl)} {\bf 63},  194 (1998);  hep-lat/9805029;
\Journal{\PRD}{61}{2000}{014015}.

\bibitem{gdecay} J. Sexton,  A.  Vaccarino  and D.  Weingarten,
{ Nucl.\ Phys.\ B (Proc.\ Suppl.)} {\bf 42} (1995)  279.

\bibitem{cmpp} C. Michael and P. Pennanen, { Phys. Rev.} {\bf D60},
(1999) 054012.

%hybrid

\bibitem{liv}  L.A. Griffiths, C. Michael and  P.E.L. Rakow,
{  Phys.\ Lett.} {\bf B129} (1983)  351.

\bibitem{pm} S. Perantonis and C. Michael, { Nucl.\ Phys.} {\bf B347}
(1990) 854.

\bibitem{jkm} K. Juge , J. Kuti and C. Morningstar, 
\Journal{\PRL}{82}{4400}{1999}; { Nucl. Phys. B (Proc. Suppl)} {\bf
83} (2000) 304,hep-lat/9909165.

\bibitem{cppacs}CP-PACS Collaboration, T. Manke et al., 
\Journal{\PRL}{82}{1999}{4396}; { Nucl. Phys. B (Proc. Suppl)} {\bf
86} (2000) 397, hep-lat/9909038; { Nucl. Phys. B (Proc. Suppl)} {\bf
83} (2000) 319, hep-lat/9909133.
     
\bibitem{milc} C. Bernard et al., { Nucl. Phys. B (Proc. Suppl.)}
{\bf 53} (1996)  228; { Phys. Rev.} {\bf D56} (1997)  7039;
{ Nucl. Phys. B (Proc. Suppl.)} {\bf 73} (1999) 264, hep-lat/9809087.

\bibitem{sesamhyb}  P. Lacock and K. Schilling,
{ Nucl. Phys. B (Proc. Suppl.)}
{\bf  73} (1999)  261,
hep-lat/9809022

\bibitem{bali} SESAM and T{$\chi$}L Collaboration, 
G. Bali et al., hep-lat/0003012.% hep-lat/9901023

\bibitem{expt}  D. Thompson et al., \Journal{\PRL}{ 79}{1997} {1630};
S. U. Chung et al., \Journal{\PRD}{60}{1999}{092001};
D. Adams et al., \Journal{\PRL}{81}{1998}{5760}

%\bibitem{cmadj}    C. Michael,   {   Nucl. Phys. B (Proc. Suppl.)}
%{\bf 6} (1992) 417.

%\bibitem{cmppsb} P. Pennanen and C. Michael, hep-lat/0001015  

\bibitem{hf8} C. Michael, { Proceedings of Heavy Flavours 8}, 
Southampton 1999,  JHEP, hep-ph/9911219.

\end{thebibliography}
\end{document}